\documentclass[submission, PhysProc]{SciPost}

\hypersetup{
    colorlinks,
    linkcolor={red!50!black},
    citecolor={blue!50!black},
    urlcolor={blue!80!black}
}

\usepackage[bitstream-charter]{mathdesign}
\DeclareSymbolFont{usualmathcal}{OMS}{cmsy}{m}{n}
\DeclareSymbolFontAlphabet{\mathcal}{usualmathcal}

  \newcommand {\nc} {\newcommand}
  \nc {\beq} {\begin{eqnarray}}
  \nc {\eeq} {\nonumber \end{eqnarray}}
  \nc {\eeqn}[1] {\label {#1} \end{eqnarray}}
  \nc {\eol} {\nonumber \\}
  \nc {\eoln}[1] {\label {#1} \\}
  \nc {\ve} [1] {\mbox{\boldmath $#1$}}
  \nc {\ves} [1] {\mbox{\boldmath ${\scriptstyle #1}$}}
  \nc {\mrm} [1] {\mathrm{#1}}
  \nc {\half} {\mbox{$\frac{1}{2}$}}
  \nc {\thal} {\mbox{$\frac{3}{2}$}}
  \nc {\fial} {\mbox{$\frac{5}{2}$}}
  \nc {\la} {\mbox{$\langle$}}
  \nc {\ra} {\mbox{$\rangle$}}
  \nc {\eq} [1] {(\ref{#1})}
  \nc {\Eq} [1] {Eq.~(\ref{#1})}
  \nc {\Ref} [1] {Ref.~\cite{#1}}
  \nc {\Refc} [2] {Refs.~\cite[#1]{#2}}
  \nc {\Sec} [1] {Sec.~\ref{#1}}
  \nc {\chap} [1] {Chapter~\ref{#1}}
  \nc {\anx} [1] {Appendix~\ref{#1}}
  \nc {\tbl} [1] {Table~\ref{#1}}
  \nc {\Fig} [1] {Fig.~\ref{#1}}
  \nc {\ex} [1] {$^{#1}$}
  \nc {\Sch} {Schr\"odinger }
  \nc {\flim} [2] {\mathop{\longrightarrow}\limits_{{#1}\rightarrow{#2}}}
  \nc {\IR} [1]{\textcolor{red}{#1}}
  \nc {\IB} [1]{\textcolor{blue}{#1}}
  \nc{\pderiv}[2]{\cfrac{\partial #1}{\partial #2}}
  \nc{\deriv}[2]{\cfrac{d#1}{d#2}}

\begin{document}

\begin{center}{\Large \textbf{
The eikonal model of reactions involving exotic nuclei;\\Roy Glauber's legacy in today’s nuclear physics\\
}}\end{center}

\begin{center}
Pierre Capel\textsuperscript{1,2$\star$}
\end{center}

\begin{center}
{\bf 1} Institut f\"ur Kernphysik, Johannes Gutenberg-Universit\"at Mainz,\\
Johann-Joachim-Becher Weg 45, D-55099 Mainz, Germany
\\
{\bf 2} Physique Nucl\'eaire et Physique Quantique (C.P. 229), Universit\'e libre de Bruxelles (ULB), 50 av. Roosevelt, B-1050 Brussels, Belgium
\\
${}^\star$ {\small \sf pcapel@uni-mainz.de}
\end{center}

\begin{center}
\today
\end{center}


\definecolor{palegray}{gray}{0.95}
\begin{center}
\colorbox{palegray}{
  \begin{tabular}{rr}
  \begin{minipage}{0.05\textwidth}
    \includegraphics[width=14mm]{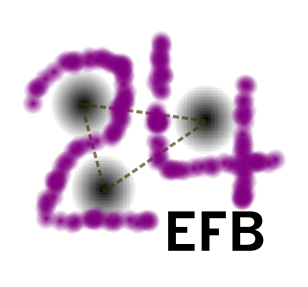}
  \end{minipage}
  &
  \begin{minipage}{0.82\textwidth}
    \begin{center}
    {\it Proceedings for the 24th edition of European Few Body Conference,}\\
    {\it Surrey, UK, 2-4 September 2019} \\
    \end{center}
  \end{minipage}
\end{tabular}
}
\end{center}

\section*{Abstract}
{\bf
In this contribution, the eikonal approximation developed by Roy Glauber to describe high-energy quantum collisions is presented.
This approximation has been---and still is---extensively used to analyse reaction measurements performed to study the structure of nuclei far from stability. 
This presentation focuses more particularly on the application of the eikonal approximation to the study of halo nuclei in modern nuclear physics.
To emphasise Roy Glauber's legacy in today's nuclear physics, recent extensions of this model are reviewed.
}

\section{Introduction}
\label{intro}

Roy J.~Glauber (1925--2018) is probably best known for his contribution to the quantum theory of optical coherence, which led the Nobel committee to award him the Nobel Prize for Physics in 2005.
He is also well known for his regular appearances at the Ig-Nobel ceremony as the ``sweeper of the broom'', where he swept the paper planes sent by the audience on the stage during the celebration\ldots

His leading role in the development of the \emph{eikonal} model of quantum collisions \cite{Glauber} is less well know by many physicists.
It is nonetheless central in the analysis of modern nuclear-physics experiments, which aim at studying the structure of nuclei far from stability (see also the contribution of Osland to this conference \cite{Osland19c}).
In particular this approximation is often used to analyse reaction measurements involving \emph{halo nuclei} \cite{Tan96}.
These very exotic systems are found at the boundaries of the nuclear chart and are characterised by a much larger matter radius than their isobars.
This unusual size is now understood as resulting from the loose binding of one or two nucleons observed in these nuclei.
Due to the purely quantal tunnel effect, these valence nucleons can be found at a large distance from the other protons and neutrons and hence form a sort of diffuse halo around a compact core \cite{HJ87}.
They can hence be described as few-body quantal systems in which one or two nucleons are loosely bound to a core, of which the structure can be neglected in first approximation.
For instance $^{11}$Be, which is one of the best known one-neutron halo nuclei, can be seen as a $^{10}$Be core in its $0^+$ ground state to which a neutron is bound by a mere 500~keV.
The even more exotic $^{11}$Li is described as a $^{9}$Li surrounded by two halo neutrons bound by 370~keV.
Although less probable, proton halos can also develop close to the proton dripline.
For example, $^8$B, in which the valence proton is bound by a mere 137~keV, is often presented as a one-proton halo nucleus.

Because they are located close to the driplines, halo nuclei are very short lived.
The half life of $^{11}$Be is, for example 13~s, while that of $^{11}$Li is less than 10~ms.
Investigating halo nuclei is done mostly through indirect techniques such as reactions \cite{Tan96}.
In particular, it has been seen that the low breakup threshold of these nuclei significantly affects their collision with other nuclei.
As shown in Refs.~\cite{Dip10,Dip12} this can already be seen in the elastic-scattering cross section.
Of course more information about the core-halo structure can be gleaned from experiments in which this structure is revealed, like in knockout and breakup reactions.
In the former, one (or two) nucleons are removed from the projectile during its interaction with a light target at high energy \cite{HT03}.
To increase the statistics of such experiments performed with rare-isotope beams, only the core is detected after the collision; the undetected halo nucleons can either be absorbed or simply scattered by the target.
This inclusive reaction provides key spectroscopic information on the single-particle structure of the nucleus \cite{Aum00}.
In breakup reactions, the halo dissociates from the core through the interaction with a target \cite{Pal03,Fuk04}.
However, contrary to the knockout reaction, this reaction is exclusive, meaning that both the core and the halo nucleon(s) are measured in coincidence after the dissociation, leading to a cleaner, although more experimentally challenging, probe.

In order to infer reliable nuclear-structure information from such reaction measurements, an accurate description of the reaction coupled to a realistic model of the projectile is needed.
The eikonal approximation developed by Glauber provides such a simple and effective model of reactions at high energy \cite{Glauber}.
In \Sec{eikonal}, after presenting the general idea behind this approximation, I will show how it can be used to describe collisions involving one-neutron halo nuclei, i.e., two-body projectiles.
Then I will describe in \Sec{reactions} the main reactions, which are used to study the halo structure at high beam energies, viz. the knockout (\Sec{KO}) and the breakup (\Sec{bu}).
Section~\ref{recent} includes recent extensions of this model: to account for relativistic effects at high energy (\Sec{rela}), to describe reactions measured on two-neutron halo nuclei, viz. with three-body projectiles (\Sec{3b}), and to use the eikonal model at low beam energies (\Sec{lowE}).
A brief summary is provided in \Sec{summary}.

\section{The eikonal approximation}\label{eikonal}
\subsection{In a nutshell}\label{nutshell}
\begin{figure}
\center
\includegraphics[width=4.cm]{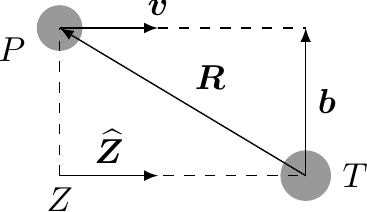}
\caption{\label{f1} System of coordinates used in the eikonal approximation to describe the collision of a one-body projectile $P$ on a structureless target $T$. The $P$-$T$ relative coordinate $\ve R$ is decomposed into its longitudinal $Z$ and transverse $\ve b$ components relative to the beam axis.}
\end{figure}

Let us start with the collision of a one-body projectile $P$ on a target $T$, which we assume structureless.
The interaction between those two bodies is simulated by an optical potential $V_{PT}$.
The \Sch equation that describes such a system reads
\beq
\left[-\frac{\hbar^2}{2\mu_{PT}}\Delta_R+V_{PT}(R)\right]\Psi(\ve{R})=E_T\Psi(\ve{R}),
\eeqn{e1}
were $\ve R$ is the $P$-$T$ relative coordinate, $\mu_{PT}$ is the $P$-$T$ reduced mass and $E_T$ is the total kinetic energy of the system in the centre-of-mass rest frame.
To describe the $P$-$T$ collision, \Eq{e1} has to be solved with the condition that initially, the projectile is far away from the target and impinging on it, i.e., that
\beq
\Psi(\ve{R})\flim{Z}{-\infty}e^{iKZ+\cdots},
\eeqn{e2}
where $K\widehat{\ve Z}$ is the wave vector for the initial $P$-$T$ relative motion, assuming $Z$ along the beam axis (see \Fig{f1}).

At sufficiently high beam energy, the wave function $\Psi$ that describes the $P$-$T$ relative motion will not differ much from the incoming plane wave of \Eq{e2}.
Indeed, under such experimental conditions, most of the reaction products will be detected at forward angles, right behind the target.
The main idea of the eikonal approximation \cite{Glauber,Osland19c} is to factorise that plane wave out of the wave function $\Psi$ to define a new wave function $\widehat\Psi$:
\beq
\Psi(\ve{R})=e^{iKZ}\ \widehat{\Psi}(\ve{R}).
\eeqn{e3}
Because the major dependence on $\ve R$ is contained in the initial plane wave, we can assume that the new wave function $\widehat\Psi$ will smoothly depend on $\ve R$.
This enables us to simplify the \Sch \Eq{e1}.
Indeed the effect of the kinetic-energy operator on $\Psi$ expressed as \eq{e3} splits into three terms
\beq
-\frac{\hbar^2}{2\mu_{PT}}\Delta_R\Psi(\ve{R})=e^{iKZ}[-\frac{\hbar^2}{2\mu_{PT}}\Delta_R -i\hbar v \frac{\partial}{\partial Z} +\frac{\mu_{PT}}{2}v^2]\ \widehat{\Psi}(\ve{R}),
\eeqn{e4}
where $v=\hbar K/\mu_{PT}$ is the initial $P$-$T$ velocity.
Because $\widehat{\Psi}$ is smoothly varying with $\ve R$, we can neglect its second-order derivative $\Delta\widehat{\Psi}$ compared to its first order derivative $K\frac{\partial}{\partial Z}\widehat{\Psi}$.
Taking into account that $E_T=\frac{\mu_{PT}}{2}v^2$, the \Sch \Eq{e1} can be rewritten as
\beq
i\hbar v \frac{\partial}{\partial Z}\widehat{\Psi}(\ve{b},Z)
=V_{PT}(R)\, \widehat{\Psi}(\ve{b},Z),
\eeqn{e5}
where $\ve R$ has been explicitly decomposed into its longitudinal $Z$ and transverse $\ve b$ components relative to the beam axis (see \Fig{f1}).
This equation is much easier to solve than \Eq{e1}: instead of a second-order differential equation depending on three variables, it is a one-order differential equation, which depends on the single longitudinal coordinate $Z$; the transverse component $\ve b$ being now a mere parameter.
The solution of \Eq{e5} that satisfies the initial condition \eq{e2} reads
\beq
\widehat{\Psi}(\ve{b},Z)&=&\exp\left[-\frac{i}{\hbar v}\int_{-\infty}^Z V_{PT}(b,Z')\,dZ'\right].
\eeqn{e6}
The $P$-$T$ wave function after the collision $\Psi(\ve{b},Z)\flim{Z}{+\infty}e^{iKZ}\ e^{i\chi(b)}$, from which the cross sections can be computed, depends on the eikonal phase
\beq
\chi(b)=-\frac{1}{\hbar v}\int_{-\infty}^\infty V_{PT}(b,Z)\,dZ.
\eeqn{e7}

The eikonal approximation therefore leads to a model of the reaction much easier to solve and which has a simple semiclassical interpretation: the projectile is seen to follow a straight-line trajectory along which it accumulates a complex phase due to its interaction with the target.
Another interest of this approximation is that it can be readily extended to two-body projectiles, such as one-nucleon halo nuclei \cite{BC12}.

\subsection{Eikonal description of reactions involving two-body projectiles}

\begin{figure}
\center
\includegraphics[width=4.5cm]{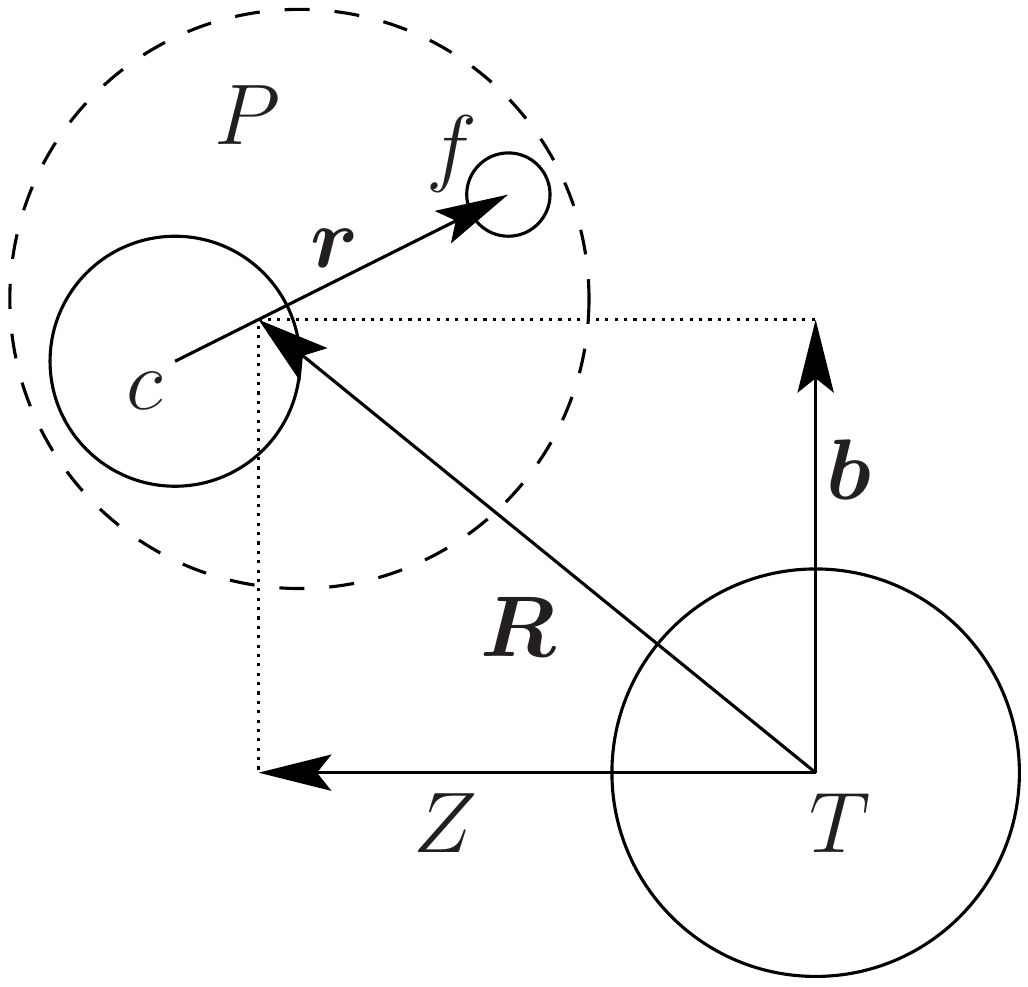}
\caption{\label{f2} Jacobi set of coordinates used to describe the collision of a two-body projectile $P$ onto a target $T$. The former is described as a fragment $f$ bound to a core $c$ separated by the internal coordinate $\ve r$.
The relative coordinate of the projectile centre of mass to the target $\ve R$ is decomposed into its longitudinal $Z$ and transverse $\ve b$ components.}
\end{figure}

To describe the collision involving a projectile $P$ that has a clear two-body structure, such as a one-nucleon halo nucleus, the internal structure of that projectile can no longer be neglected.
Modelling $P$ as a core $c$ to which a fragment $f$ is loosely bound, its structure is usually described by the single-particle quantum-mechanical Hamiltonian
\beq
H_0=-\frac{\hbar^2}{2\mu}\Delta_r+V_{cf}(r),
\eeqn{e8}
where $\ve r$ is the $c$-$f$ relative coordinate (see \Fig{f2}), $\mu$ is the reduced mass of the projectile constituents and $V_{cf}$ is an effective potential whose parameters are adjusted to reproduce the bound states of the projectile and, sometimes, some of its low-lying resonances \cite{BC12,CGB04}.

The internal structure of the target $T$ is usually neglected and its interaction with the projectile constituents is simulated by optical potentials chosen in the literature.
Within that framework, studying the $P$-$T$ collision reduces to solving the three-body \Sch equation
\beq
\left[-\frac{\hbar^2}{2\mu_{PT}}\Delta_R+H_0+V_{cT}(R_{cT})+V_{fT}(R_{fT})\right]\Psi(\ve{r},\ve{R})=E_T\Psi(\ve{r},\ve{R}),
\eeqn{e9}
where $\ve{R}$ is the coordinate of the projectile centre of mass relative to the target (see \Fig{f2}), and $V_{cT}$ and $V_{fT}$ are the $c$-$T$ and $f$-$T$ optical potentials, which depend on the $c$-$T$ and $f$-$T$ relative distances, respectively.
This equation has to be solved with the condition that the projectile is initially in its ground state
\beq
\Psi(\ve{r},\ve{R})\flim{Z}{-\infty}e^{iKZ+\cdots}\ \Phi_0(\ve{r}),
\eeqn{e10}
where $\Phi_0$ is the eigenstate of the Hamiltonian $H_0$ \eq{e8} of eigenenergy $\epsilon_0$, which corresponds to the bound state of the projectile.
As before, the $Z$ axis is chosen along the incoming beam and the $P$-$T$ wave vector $K\widehat{\ve Z}$ is related to the total energy of the system $E_T=\frac{\hbar^2}{2\mu_{PT}}K^2+\epsilon_0$.

In this case, the idea of the eikonal approximation is similar to that in the case of a one-body projectile: the incoming plane wave $e^{iKZ}$ is factorised out of the three-body wave function $\Psi$ to lead to a simpler equation to solve for the new wave function $\widehat\Psi$ [see \Eq{e3}] \cite{BC12,BCG05}:
\beq
i\hbar v \frac{\partial}{\partial Z}\widehat{\Psi}(\ve{r},\ve{b},Z)
=[H_0-\epsilon_0+V_{cT}(R_{cf})+V_{fT}(R_{fT})]\,
\widehat{\Psi}(\ve{r},\ve{b},Z),
\eeqn{e11}
which needs to be solved for each value of $\ve b$, the transverse component of $\ve R$ with the initial condition
\beq
\widehat\Psi(\ve{r},\ve{b},Z)\flim{Z}{-\infty}\Phi_0(\ve{r}).
\eeqn{e12}
Because the reactions described in the eikonal approximation must take place at high energy, it is usual to perform an additional adiabatic---or sudden---approximation, which neglects the excitation energy of the projectile compared to the beam energy: $H_0-\epsilon_0\approx0$.
In that case, the eikonal solution of \Eq{e9}, which satisfies the initial condition \Eq{e10}, reads \cite{BC12,HT03}
\beq
\Psi(\ve{r},\ve{b},Z)&\flim{Z}{+\infty}&e^{iKZ}\ \exp\left\{i\left[\chi_{cT}(\ve{r},\ve{b})+\chi_{fT}(\ve{r},\ve{b})\right]\right\}\Phi_0(\ve{r}),
\eeqn{e13}
in which the eikonal phases $\chi_{cT}$ and $\chi_{fT}$ are computed following \Eq{e7} for the $c$-$T$ and $f$-$T$ optical potentials, respectively.

Thanks to its simplicity, the eikonal approximation has been widely used to study the spectroscopy of halo nuclei through reactions and mostly knockout reactions \cite{HT03}, which is performed at high energy and on light targets, for which this model of reactions is well suited.

\section{Reactions with halo nuclei}\label{reactions}
\subsection{Knockout}\label{KO}

In knockout reactions, a nucleon is quickly removed from the projectile during a high-energy collision measured on a light target (usually C or Be).
It is one of the ways to obtain spectroscopic information on exotic nuclei \cite{HT03}.
In particular, it has been extensively used to probe the one-neutron halo structure \cite{Aum00,Sau00}.
In that case, the removed neutron either breaks up from the core of the nucleus, due to its low separation energy, or it is absorbed by the target.
In such experiments, only the core is measured as a product of the reaction, which requires less high beam intensity than if the halo neutron were to be measured in coincidence, as is the case in breakup reactions (see \Sec{bu}).

Because the reaction is performed at high energy and on a light target, it can be seen as taking place adiabatically---viz. suddenly---which means that the detected core will retain some information about its state in the initial ground state of the projectile.
Since the halo wave function is characterised by a large spatial expansion, we can expect the core to have a fairly narrow momentum distribution within the projectile.
Hence the idea to measure the core momentum distribution after knockout to infer spectroscopic information about halo nuclei \cite{HT03}.
A narrow distribution would be a strong indication that the nucleus exhibits a one-neutron halo.
Extensive campaigns of measurements have been organised to measure the momentum distribution along isotopic lines, as the one reported in \Ref{Sau00}.

\begin{figure}
\center
\includegraphics[page=3, trim=11.7cm 3.8cm 2.5cm 17cm, clip=true, width=7.cm]{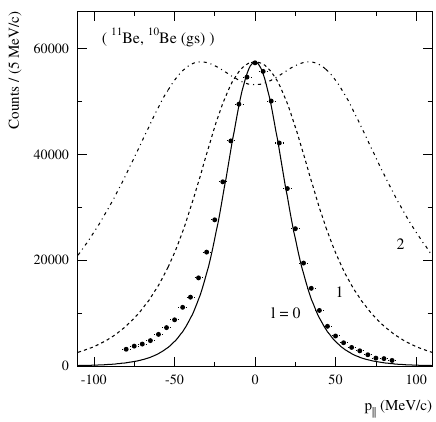}
\caption{\label{f3} Parallel-momentum distribution of the $^{10}$Be core of the one-neutron halo nucleus $^{11}$Be following its one-neutron removal reaction on Be at 60~MeV/nucleon.
Eikonal calculations performed assuming different orbital-angular momenta of the halo neutron $l$ are compared to the experimental data of \Ref{Aum00}, illustrating the spectroscopic application of this reaction observable in the study of exotic nuclei.
Reprinted figure with permission from \Ref{Aum00}. Copyright (2000) by the
American Physical Society.}
\end{figure}

Such a typical measurement is illustrated in \Fig{f3}, which displays the parallel-momentum distribution of $^{10}$Be following the removal of one neutron from $^{11}$Be measured at 60~MeV/nucle\-on on a Be target \cite{Aum00}.
In addition to the experimental data, three calculations performed within the usual eikonal approximation are also shown.
Each one of them assumes a different orbital angular momentum for the valence neutron in this archetypical one-neutron halo nucleus: $l=0$ (solid line), 1 (dashed line) and 2 (dash-dotted line).
We observe that the first of these, which agrees with the data, leads to the narrowest distribution, which is to be expected in this case where no barrier hinders the development of a long-range halo in the projectile wave function.
Not only can this type of measurement hint at possible halo nuclei, but it can also be used as a spectroscopic tool to pinpoint to single-particle orbital in which the valence nucleon sits.

\subsection{Breakup}\label{bu}
In breakup reactions, the halo nucleon(s) dissociate from the core during the collision with the target.
However, contrary to knockout, where only the resulting core is measured, all constituents, i.e., the core and the halo neutron(s), are measured in coincidence.
As mentioned before, this requires higher beam intensities than for knockout experiments since the measurement of neutrons is not an easy task.
However, this reaction provides more detailed information, in particular on the structure of the low-energy continuum of the projectile \cite{Fuk04,CGB04}.

To infer reliable information about the structure of the projectile from experiment, an accurate model of this reaction is needed.
Unfortunately, in its most usual expression, the eikonal approximation is not valid to describe reactions on heavy targets.
In these cases, the breakup is dominated by the Coulomb interaction, which, due to its infinite range, is incompatible with the adiabatic approximation that assumes that the reaction takes place on a very short time.
Mathematically this translates in an eikonal phase that behaves as $\frac{1}{b}$, leading to a divergence in the integral over $b$ in the calculation of the cross section \cite{GBC06}.

\begin{figure}
\center
\includegraphics[width=7.3cm]{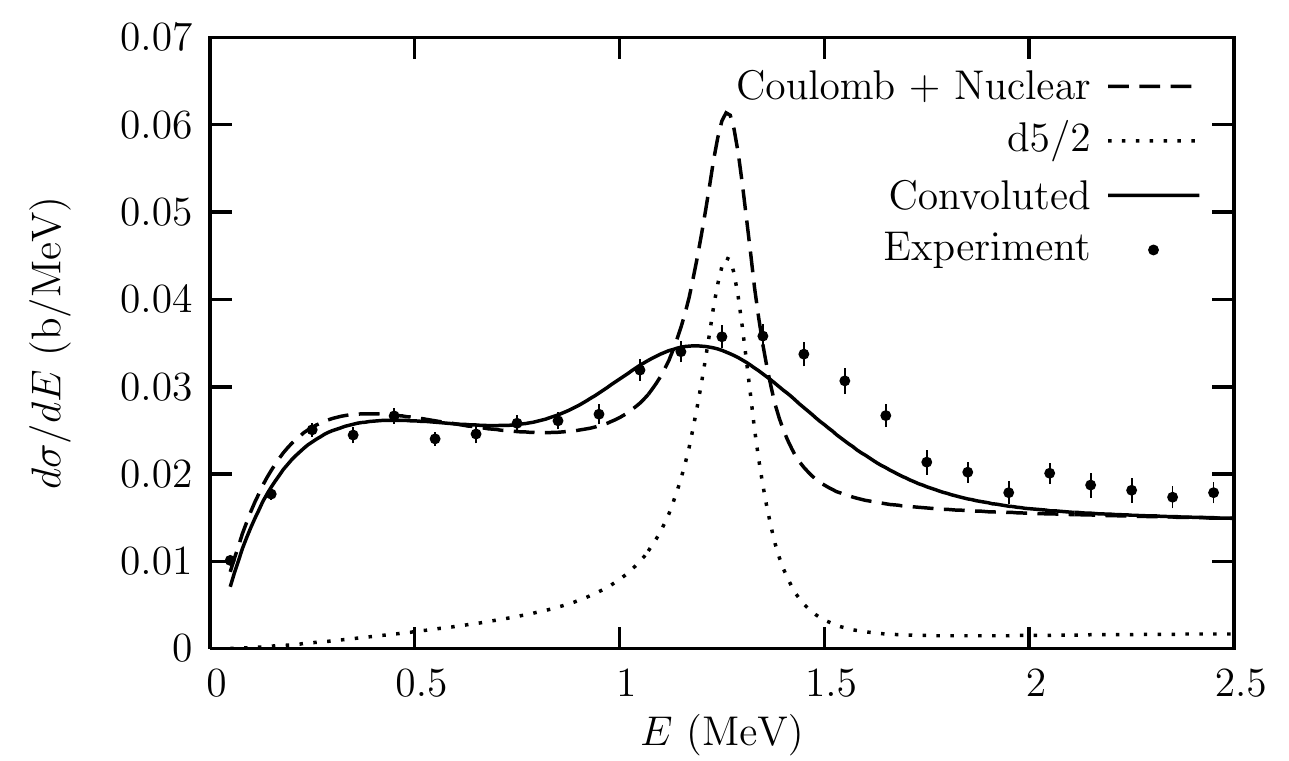}
\includegraphics[width=7.cm]{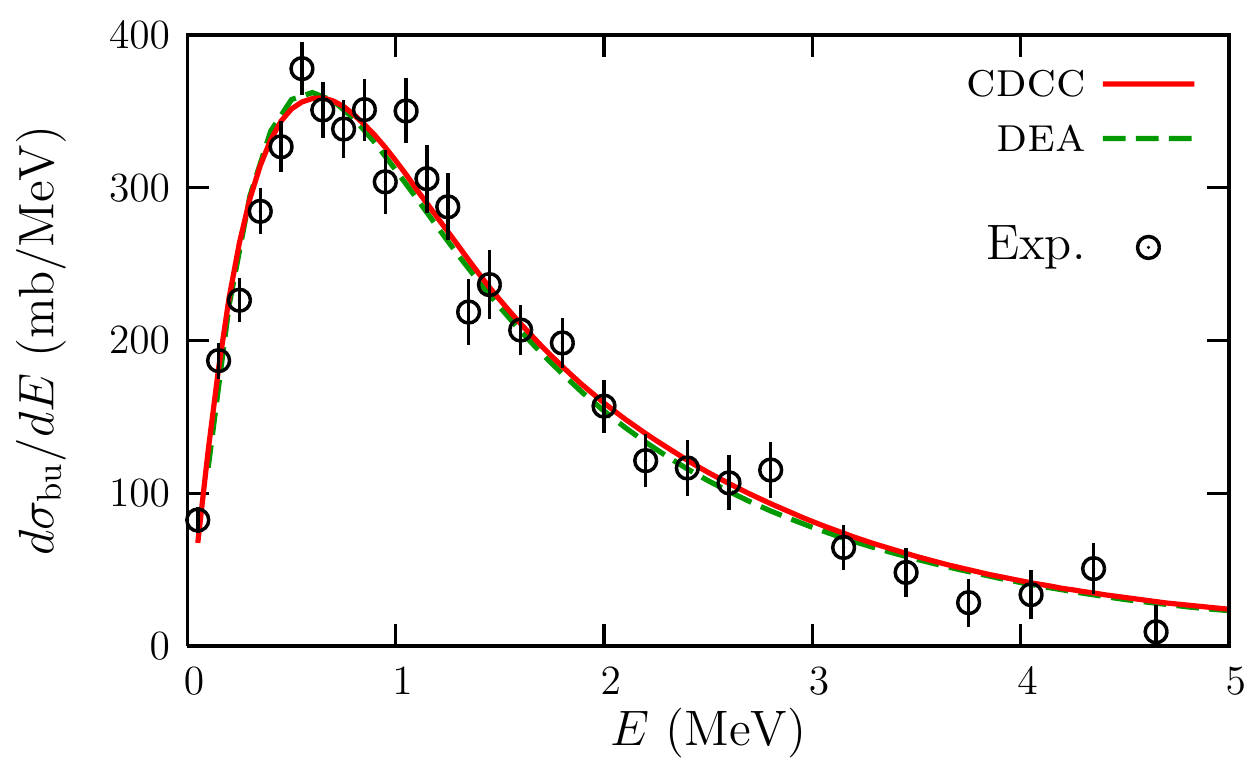}
\caption{\label{f4} Cross sections for the breakup of (left) $^{11}$Be on C at 67~MeV/nucleon and (right) $^{15}$C on Pb at 68~MeV/nucleon, both expressed as a function of the $c$-n relative energy $E$ after dissociation.
Dynamical eikonal calculations \cite{GBC06,CEN12} are compared to the experimental data of (left) \Ref{Fuk04} and (right) \Ref{Nak09}.
The good agreement with experiment confirms the few-body structure of the one-neutron halo projectiles.
Right figure adapted with permission from \Ref{CEN12}. Copyright (2010) by the
American Physical Society.}
\end{figure}

To avoid this divergence, it has been suggested not to perform the adiabatic approximation.
This requires to solve \Eq{e11} numerically.
This Dynamical Eikonal Approximation (DEA) \cite{BCG05} has shown to work very well to describe various reaction observables on both light and heavy targets at intermediate energy for one-neutron \cite{GBC06} and one-proton halo nuclei \cite{GCB07}.
These results are illustrated in \Fig{f4}, where the breakup cross section of one-neutron halo nuclei $^{11}$Be (left) and $^{15}$C (right) are plotted as a function of the relative energy $E$ between the neutron and the core after dissociation.

The reaction involving $^{11}$Be shown in \Fig{f4}~(left) has been measured on C at 67~MeV/nucleon \cite{Fuk04}.
The cross section exhibits a significant peak at about $E=1.3$~MeV, which corresponds to the energy of a known $\fial^+$ resonant state in the spectrum of $^{11}$Be, suggesting that information about the projectile continuum can be gained through such measurements.
The DEA calculation confirms this.
When including that resonances within the $d_{5/2}$ partial wave, a narrow peak appears in that contribution to the cross section (dotted line).
Once folded with experimental resolution (solid line), the theoretical prediction of \Ref{CGB04} agrees very well with the data of \Ref{Fuk04}.

The breakup of $^{15}$C has been measured on a lead target at 68~MeV/nucleon (right panel of \Fig{f4}) \cite{Nak09}.
The DEA calculation (dashed green line) is in perfect agreement with experiment \cite{CEN12}, confirming the clear two-body structure of $^{15}$C.
Interestingly, this result is also in perfect agreement with a similar calculation performed within the Continuum Discretised Couple Channel model (CDCC, red solid line), which is fully quantal and does not rely on any approximation about the projectile-target relative motion \cite{BC12,Aus87}.
This comparison confirms the validity of the eikonal approach at intermediate energy \cite{CEN12}.

It should be noted that although much less time-consuming than CDCC, the DEA is still much more computationally expensive than the usual eikonal approximation, which requires simple one-dimensional integrals to account for the $P$-$T$ interactions [see \Eq{e7}].
To cure the aforementioned divergence due to the Coulomb interaction, it has been suggested to replace the erroneous Coulomb eikonal phase at by its (correct) estimate at the first order of the perturbation theory \cite{MBB03}.
This Coulomb-Corrected Eikonal model (CCE) has been shown to agree with DEA calculations at intermediate energy and hence to provide a simple and computationally cheap alternative to the more sophisticated dynamical models \cite{CBS08}.
Note that a more advanced correction has been recently suggested following a similar idea \cite{HB19a}.

\section{Recent extensions of the eikonal approximation}\label{recent}
\subsection{Relativistic beam energies}\label{rela}
The breakup measurements performed at GSI, as well as those now made at RIKEN, involve very high-energy beams, of the order of a few hundreds MeV/nucleon.
At these energies it is clear that a proper treatment of special relativity is needed.
Since these energies are perfectly suited for the adiabatic approximation, but for the Coulomb interaction, the aforementioned CCE \cite{MBB03,CBS08} is a good starting point for the development of such a model.
Recently, it has been shown that treating the nuclear interaction at the Optical Limit Approximation (OLA) of the Glauber model \cite{Glauber,BD04}, and correcting for relativistic effects the sole treatment of the Coulomb interaction in a CCE model, a good agreement with the experimental breakup cross section of $^{11}$Be on Pb measured at 520~MeV/nucleon \cite{Pal03} could be obtained \cite{MC19}.

\begin{figure}
\center
\includegraphics[width=7.3cm]{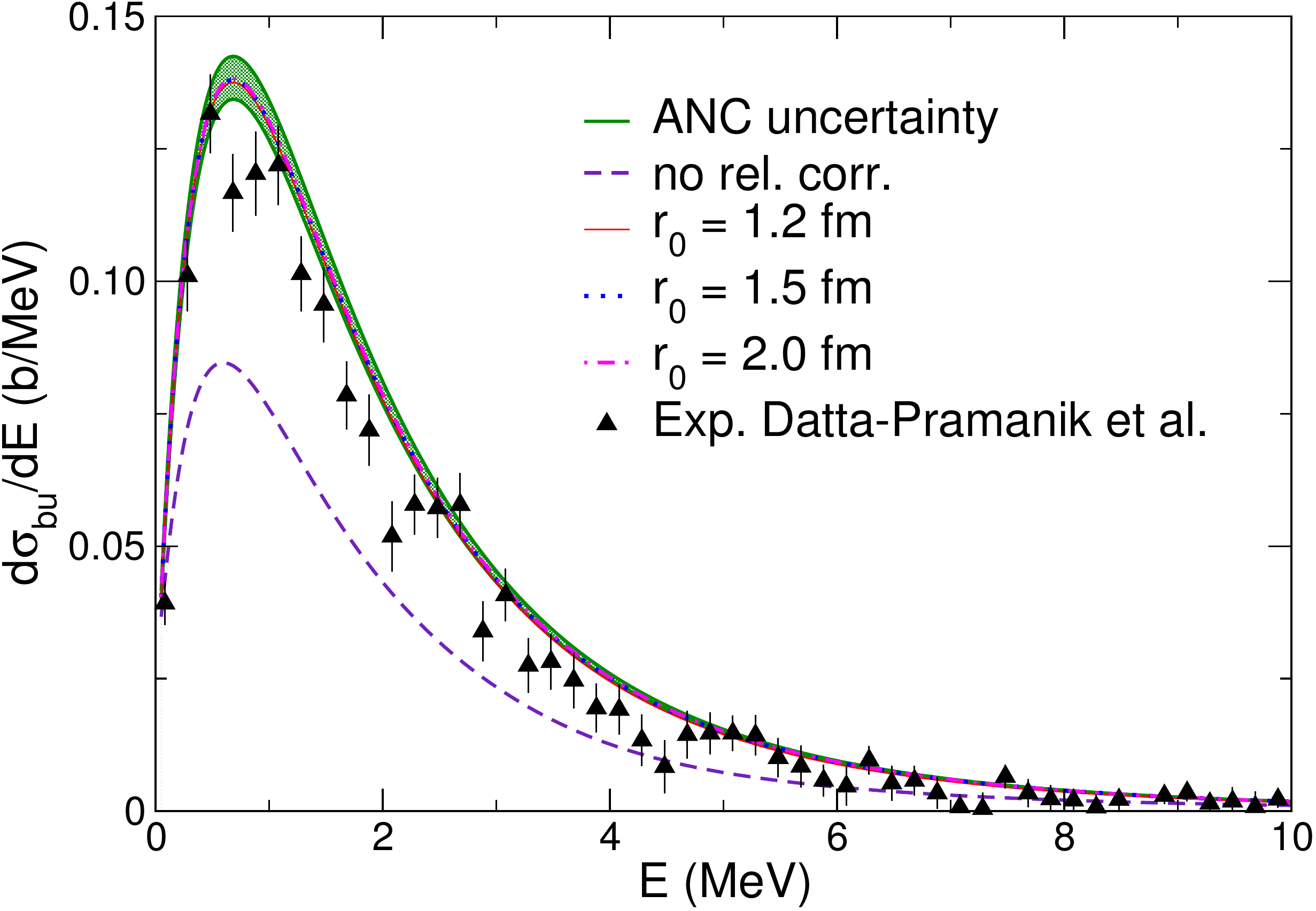}
\caption{\label{f5} Cross sections for the breakup of $^{15}$C on Pb at 605~MeV/nucleon expressed as a function of the $c$-n relative energy $E$ after dissociation.
Coulomb-corrected eikonal calculation \cite{MYC19} are compared to the experimental data of \Ref{Pra03}.
Reprinted figure with permission from \Ref{MYC19}. Copyright (2019) by the
American Physical Society.}
\end{figure}

Figure~\ref{f5} illustrates this new model applied to the Coulomb breakup of $^{15}$C, measured at GSI at 605~MeV/nucleon \cite{Pra03}.
The results of calculations using different models of $^{15}$C are in excellent agreement with the data (red solid, blue dotted and magenta dash-dotted lines) \cite{MYC19}.
The purple dashed line, which corresponds to the result of a calculation without relativistic correction, confirms the significance of relativistic effects and hence the necessity to include them properly within the description of the reaction.

\subsection{Three-body projectiles}\label{3b}
The computational efficiency and reliability of the CCE also opens the door to a more complex description of the projectile.
In particular it has been proposed to extend this model to three-body projectiles, viz. two-neutron halo nuclei \cite{BCD09,BC12}.

\begin{figure}
\center
\includegraphics[width=7.5cm]{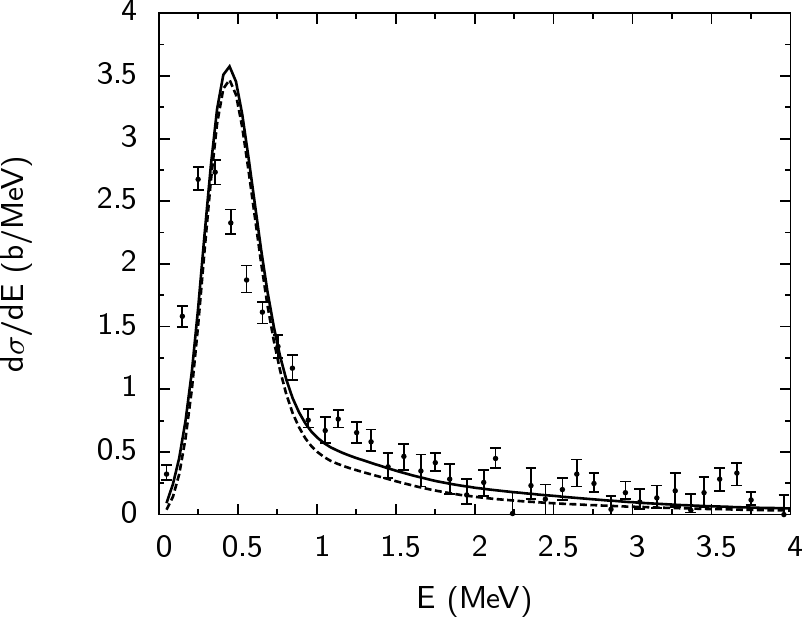}
\includegraphics[width=7cm]{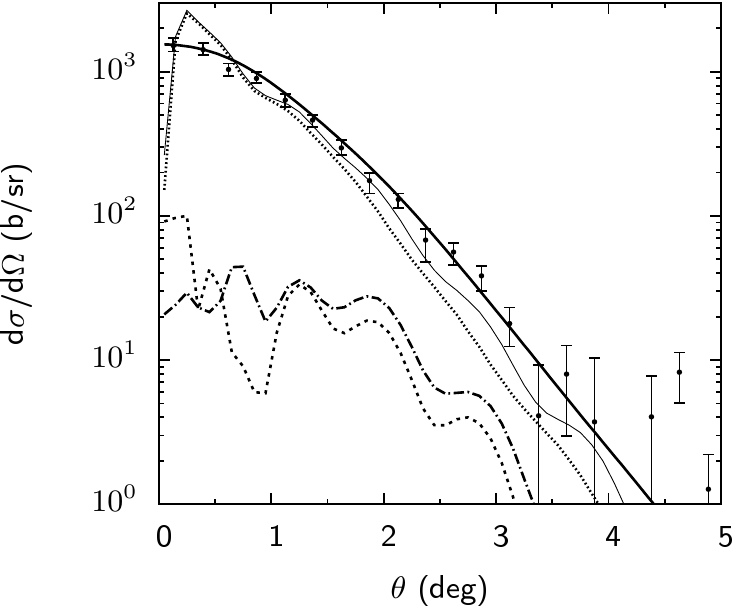}
\caption{\label{f6} Cross sections for the breakup of $^{11}$Li on Pb at 70~MeV/nucleon expressed as a function of (left) the $^9$Li-n-n relative energy $E$ after dissociation and (right) the $^9$Li-n-n centre of mass.
Coulomb-corrected eikonal calculations \cite{PDB12} are compared to the experimental data of \Ref{Nak06}.
Reprinted figure with permission from \Ref{PDB12}. Copyright (2012) by the
American Physical Society.}
\end{figure}

Figure~\ref{f6} illustrates the excellent results obtained with this model for $^{11}$Li\cite{PDB12}, the archetypical two-neutron halo nucleus whose Coulomb breakup has been accurately measured on Pb at 70~MeV/nucleon \cite{Nak06}.
The good agreement with experiment observed on both the energy (left panel of \Fig{f6}) and angular (right panel of \Fig{f6}) distributions confirms both the validity of the three-body description of $^{11}$Li used in \Ref{PDB12} and the possibility to extend the CCE to the study of Borromean nuclei.

\subsection{Low-energy reactions}\label{lowE}
As illustrated so far, the eikonal approximation is very reliable to describe reactions measured at intermediate and high energies, viz. above 50~MeV/nucleon.
However, the computing attractiveness of this approximation over more quantal models, like CDCC, has raise the interest to find a correction to extend its range of validity to lower beam energies.
In particular, an eikonal-like model valid at about 10~MeV/nucleon would be valuable to analyse data taken at the new HIE-ISOLDE facility at CERN or at the future ReA12 of FRIB.

\begin{figure}
\center
\includegraphics[width=7.3cm]{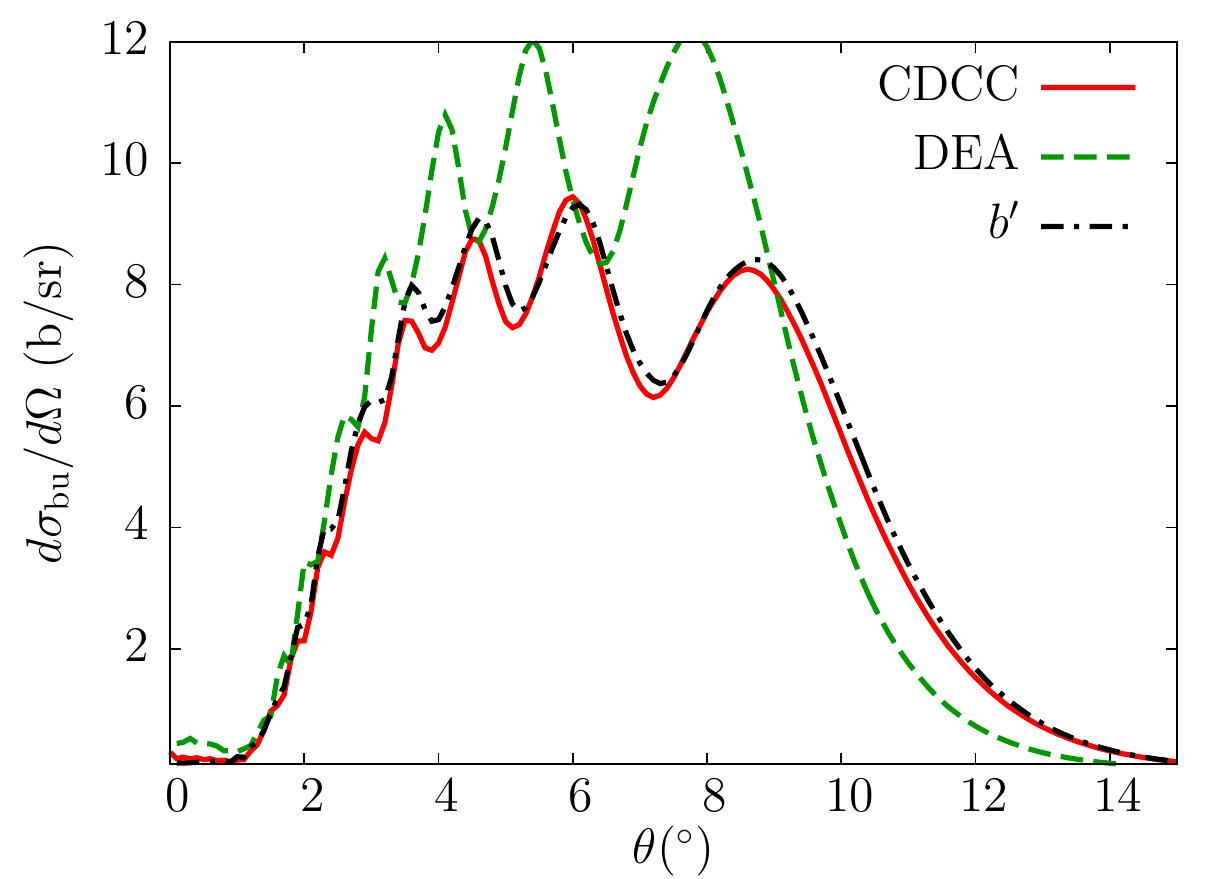}
\caption{\label{f7} Angular distribution for the breakup of $^{15}$C on Pb at 20~MeV/nucleon.
Dynamical eikonal calculations and its low-energy correction are compared fully quantal CDCC results \cite{FOC14}.
The simple low-energy correction enables to reproduce the correct magnitude and oscillatory pattern of the cross section.
Figure adapted from \Ref{FOC14}. Copyright (2014) by the
American Physical Society.}
\end{figure}

In the detailed comparison between the DEA and CDCC presented in \Ref{CEN12}, it has been shown that already at 20~MeV/nucleon, the eikonal approximation fails at reproducing CDCC cross sections (see \Fig{f7}).
It appears that the DEA (dashed green line) leads to too large and too forward-focussed an angular distribution compared to the CDCC reference calculation (red solid line).
This can be easily understood from the semiclassical interpretation of the eikonal approximation (see \Sec{nutshell}).
The projectile is seen as following a straight-line trajectory.
It is thus ``forced'' towards forward angles and constrained to pass through the high-field zone of the target.
It is thus not surprising that the resulting cross section appears too high and peaking at too forward an angle, simply because the Coulomb deflection is not included in the model.

In a later analysis, it has been shown that a simple semiclassical correction is enough to account for the Coulomb deflection \cite{FOC14}.
The idea of that correction is to replace the transverse coordinate $b$ of $\ve R$ in the DEA \Eq{e11} by the classical distance of closest approach $b'$ for the Coulomb trajectory corresponding to the impact parameter $b$ \cite{BD04}
\beq
b'=\frac{\eta}{K}+\sqrt{\frac{\eta^2}{K^2}+b^2},
\eeqn{e14}
where the Sommerfeld parameter $\eta=Z_PZ_Te^2/4\pi\epsilon_0\hbar v$, and $Z_P$ and $Z_T$ are the projectile and target atomic numbers, respectively.
With this simple correction, the DEA cross section is reduced and extended to larger angles and perfectly matches the CDCC prediction (see the black dash-dotted line in \Fig{f7}) \cite{FOC14}.

At least for Coulomb-dominated collisions it is thus possible to extend the eikonal approximation to low energies.
It has been tried to apply similar corrections to nuclear-dominated reactions, hence to light targets \cite{AZV97,BAT99}.
Unfortunately, none of these corrections seem to work to properly describe the breakup of one-neutron halo nuclei on light targets down to 20~MeV/nucleon \cite{HC18}.
Although disappointing, this result shows the limit of the eikonal approximation and in which conditions it is requested to resort to fully quantal models.




\section{Summary}\label{summary}
Roy Glauber has developed the eikonal approximation to describe high-energy quantal collisions \cite{Glauber}.
At least in the realm of nuclear physics, his model has proved---and continues to prove---very helpful to accurately analyse experiments performed at radioactive-ion beam facilities to study the structure of nuclei away from stability \cite{Osland19c,BC12}.

In this contribution, I have illustrated the use of the eikonal approximation to analyse reactions involving halo nuclei \cite{Tan96}.
These exotic, short-lived nuclei exhibit a very unusual structure, where one or two loosely-bound nucleons, tunneling into the classically-forbidden region, exhibit a high probability of presence at a large distance from the other nucleons.
These valence nucleons hence form a sort of diffuse halo around the compact core of the nucleus \cite{HJ87}.
This peculiar structure challenges the usual description of nuclei, in which nuclei are seen as compact clusters of nucleons.
Because of their short lifetime, halo nuclei are often studied through reactions.
To infer reliable spectroscopic information from such measurements, a good understanding of the reaction process is needed.
The eikonal approximation provides us with such a model for reactions measured at high beam energy (viz. above 50~MeV/nucleon).

In particular, I have emphasised the use of the usual eikonal approximation, i.e., which includes the subsequent adiabatic treatment of the projectile dynamics, to analyse measurements of knockout reactions \cite{HT03}, i.e., of reactions where the halo nucleons are removed from the nucleus on a light target at high energy, see, e.g., \Ref{Aum00}.

The use of the eikonal approximation to describe breakup reactions, in which both the core and the halo nucleons are measured in coincidence \cite{Pal03,Fuk04}, requires a correct treatment of the $P$-$T$ Coulomb interaction, which cannot be accounted for at the adiabatic approximation \cite{BC12,BCG05,CBS08}.
This can be done either by fully including the dynamics of the projectile within the reaction model, like in the DEA \cite{BCG05,GBC06} or by correcting the eikonal treatment of the Coulomb interaction at the first order of the perturbation theory \cite{MBB03,CBS08,HB19a}.

The eikonal approximation provides a simple and elegant model of the reaction with a clear semiclassical interpretation.
Unfortunately it is limited to high beam energies, viz. above 50~MeV/nucleon.
Efforts have been made to extend its range of validity towards lower energies.
For reactions on heavy targets, the Coulomb deflection can be easily simulated by a simple semiclassical correction, enabling a correct description of breakup reactions down to, at least, 20~MeV/nucleon \cite{FOC14}.
On light targets however, no such correction seems to provide acceptable results to reliably analyse reactions at low energy \cite{HC18}.

The eikonal approximation and the models that have been derived have provided and will continue to provide valuable information on the structure of nuclei away from stability, such as halo nuclei.
This confirms the significant, albeit less known, contribution of Roy Glauber in modern nuclear physics.



\paragraph{Funding information}
This project has received funding from the European Union’s Horizon 2020 research and innovation programme under Grant Agreement No. 654002, the Deutsche Forschungsgemeinschaft within the Collaborative Research Centers 1044 and 1245, and the PRISMA+ (Precision Physics, Fundamental Interactions and Structure of Matter) Cluster of Excellence.
I acknowledge the support of the State of Rhineland-Palatinate.

\nolinenumbers

\end{document}